# A Biomedically oriented automatically annotated Twitter COVID-19 Dataset

Authors: Luis Alberto Robles Hernandez[1], Tiffany J. Callahan[2], Juan M. Banda[1]*

[1] Department of Computer Science, Georgia State University, Atlanta, Georgia, 30303 USA
[2] Computational Bioscience Program, University of Colorado Anschutz Medical Campus, Aurora, CO, 80045 USA
* Corresponding author: jmbanda@gsu.edu

## Abstract:

The use of social media data, like Twitter, for biomedical research has been gradually increasing over the years. With the COVID-19 pandemic, researchers have turned to more non-traditional sources of clinical data to characterize the disease in near-real time, study the societal implications of interventions, as well as the sequelae that recovered COVID-19 cases present (Long-). However, manually curated social media datasets are difficult to come by due to the expensive costs of manual annotation and the efforts needed to identify the correct texts. When datasets are available, they are usually very small and their annotations don't generalize well over time or to larger sets of documents. As part of the 2021 Biomedical Linked Annotation Hackathon, we release our dataset of over 120 million automatically annotated tweets for biomedical research purposes. Incorporating best-practices, we identify tweets with potentially high clinical relevance. We evaluated our work by comparing several SpaCy-based annotation frameworks against a manually annotated gold-standard dataset. Selecting the best method to use for automatic annotation, we then annotated 120 million tweets and released them publicly for future downstream usage within the biomedical domain.

## Introduction

Social media platforms like Twitter, Instagram, and Facebook provide researchers with unprecedented insight into personal behavior on a global scale. Twitter is currently one of the leading social networking services with over 353 million users and reaching ~6% of the world's population over the age of 13 [1]. It is also quickly becoming one of the most popular platforms for conducting health-related research because of its use for public health surveillance, pharmacovigilance, event detection/forecasting, and disease tracking [2,3]. During the last decade, Twitter has provided substantial aid in the surveillance of pandemics, including the Zika virus [4], H1N1 (or Swine Flu) [5], H7N9 (or avian/bird flu) [6], and Ebola

[7]. Twitter has been used extensively during the 2020 COVID-19 outbreak [8], providing insight into everything from monitoring communication between public health officials and world leaders [9], tracking emerging symptoms [10] and access to testing facilities [11], to understanding the public's top fears and concerns about infection rates and vaccination [12]. While it is clear that Twitter contains invaluable content that can be used for a myriad of benevolent endeavors, there are many challenges to accessing and leveraging these data for clinical research and/or applications.

Researchers face a myriad of challenges when trying to utilize Twitter data. Aside from the potential ethical challenges, which will not be discussed in this work (see [13] for a review of this area), it can be difficult to obtain access to these data and hard to keep up with real-time content collection [14,15]. Once the data have been obtained, researchers must then perform several preprocessing steps to ensure the data are sufficient for analysis. Concerning COVID-19, there are several existing social media repositories [16–20]. Unfortunately, most of these repositories are infrequently updated, do not provide any preprocessing or data cleaning, and either do not provide the raw data or lack appropriate metadata or provenance. The COVID-19 Twitter Chatter dataset [20] is a robust large-scale repository of tweets that is well-maintained and frequently updated (over 50 versions released at the time of publication). Recent work utilizing this resource has shown great promise for tracking long-term patient-reported symptoms [21] as well as highlighted mentions of drugs relevant to the treatment of COVID-19 [22]. While these are compelling clinical use cases, additional work is needed to fully understand what additional biomedical and clinical utility can be obtained from these data.

This paper presents preliminary work achieved during the 2021 Biomedical Linked Annotation Hackathon (BLAH 7) [23], which aimed to enhance and extend the COVID-19 Twitter Chatter dataset [20] to include biomedical entities. By annotating symptoms and other relevant biomedical entities from COVID-19 tweets, we hope to improve the downstream clinical utility of these data and provide researchers with a means to clinically characterize personally-reported COVID-19 phenomena. We envision this work as the first step towards our larger goal of deriving mechanistic insights from specific types of entities within COVID-19 tweets by integrating these data with larger and more complex sources of biomedical knowledge, like PheKnowLator [24] and the KG-COVID-19 [25] knowledge graphs. The remainder of this paper is organized as follows: an overview of the methods and technologies utilized in this work, an overview of our findings, and a brief discussion of conclusions and future work.

# Methods

To prepare the dataset released in this work, we looked for named entity recognition (NER) pipelines to identify biomedical entities in text. We opted to evaluate: MedSpaCy [26], MedaCy [27], and ScispaCy [28], alongside a traditional text annotation pipeline from Social Media Mining Toolkit (SMMT), a product of a BLAH 6 hackathon [29]. The main reason for selecting these text processing pipelines is the fact that they are all based on Spacy [30], a widely adopted open-source library for Natural Language Processing (NLP) in Python, allowing our codebases to be streamlined, and the annotation output to be easily compared in our evaluation as well as ingested by other work utilizing similar pipelines. Several preprocessing steps like URL and emoji removal were performed on all tweets.

Please note that the selected NER pipelines are usually tuned and developed to annotate specific types of clinical/scientific text, from either electronic health records, clinical notes, or scientific literature. The only general-purpose tagger is the Social Media Mining Toolkit (SMMT), which does not perform any specialized tasks other than tagging or annotating text. This fact impacted their performance in Twitter social media data, and the following comparison should not be used to evaluate the systems' performance on clinical data/scientific literature, but rather the need for appropriately tuned systems for social media data.

## Datasets

As the source for this work, we used one of the largest COVID-19 Twitter chatter datasets available [20]. We used version 44 of the dataset [20], which contains 903,223,501 unique tweets. To improve the quality and relevance of the annotations, we used the clean version of this dataset, which has all retweets removed. Leaving us with a total of 226,582,903 unique tweets to annotate. From this subset, we selected only English tweets, as all the systems evaluated were created to extract/annotate biomedical concepts in this language.

For the evaluation of the annotations from each NER system and the SMMT tagger, we will use as a gold standard, a manually annotated dataset created for symptoms, conditions, prescriptions, and measurement procedures identification in patients with long Covid phenotypes [21]. This dataset consists of 10,315 manually annotated tweets, by multiple clinicians. Currently, the dataset is not publicly available but will be released at a later date.

## ScispaCy

Developed by the Allen AI institute, the pipelines and models in this package have been tuned for use on scientific documents [28]. In our evaluation, we used the following model: *en_core_sci_lg*, which consists of ~785k vocabulary and 600k word vectors. Additionally, we used the EntityLinker component to annotate the Unified Medical Language System (UMLS) concepts. Since this pipeline provides more than one match per annotation, we only selected the first match to avoid duplicates. The code used can be found in [31]

## MedaCy

Developed by researchers at Virginia Commonwealth University, MedaCy is a text processing framework wrapper for spaCy. It supports extremely fast prototyping of highly predictive medical NLP models. For our evaluation, we used their provided *medacy_model_clinical_notes* model, with all other default settings. The code used can be found in [31]

## MedSpaCy

Currently, in beta release, MedSpaCy was created as a toolkit to enable user-specific clinical NLP pipelines. In our evaluation, we wanted to use some of the out-of-the-box components instead of fine-tuning them for our Twitter annotation task. We used the *en_info_3700_i2b2_2012* model - trained on i2b2 data, and the Sectionizer ([32]). We initially tried to use the demo QuickUMLS entity linker, but ultimately opted not to do this as their demo only includes 100 concepts, and building it from scratch was outside of the scope of our task. The code used can be found in [31]

## SMMT Tagger

As part of SMMT, the spaCy-based tagger relies on a user-specified dictionary to annotate concepts on the provided text. This tagger does not perform any NER or section detection, but only simple string matching. Designed with simplicity and flexibility in mind, when using social media data it is preferred to provide a concise dictionary with the desired terms for annotation, rather than using pre-trained models that may not generalize well to domain-specific tasks, or are computationally expensive. The dictionary used in this evaluation consists of a mix of SNOMED-CT [33], ICD 9/10 [34], MeSH [35], and RxNorm [36] extracted from the Observational Health Data Sciences and Informatics (OHDSI) vocabulary. This dictionary is available as part of the paper's code repository.

# Results

## Extraction performance

In Table 1 we show the processing time and count of annotations produced by the evaluated systems on the gold standard dataset. Note that as expected, simple text annotation from the SMMT tagger is the fastest, with MedaCy coming in second as its annotation model is small. The SMMT tagger dictionary produces plenty of annotations as it considers some of the common misspellings for COVID-19 (e.g. "fatigue" vs "fatige") as well as related symptoms and drugs that have been curated in our previous work when extracting drug mentions in Twitter data [22].

|  | **Tweets** | **Annotations Produced** | **Processing Time (seconds)** |
|---|---|---|---|
| **SMMT Tagger** | 10,315 | 92,835 | 10,815.24 |
| **MedSpaCy** | 10,315 | 51,575 | 33,746.40 |
| **MedaCy** | 10,315 | 61,890 | 21,896.63 |
| **ScispaCy** | 10,315 | 72,205 | 49,168.85 |

Table 1. Extraction evaluation of proposed systems.

Due to the larger model utilized by ScispaCy, the processing time is nearly five-fold that of simple text annotation. However, this comes with the added benefit that abbreviations are nicely normalized to UMLS concepts, hence creating some annotations that any of the other systems will be unable to find.

## Overlap between systems on gold standard

To determine which system to use for the large-scale annotation of the Twitter COVID-19 chatter dataset, we evaluated all systems against the manually annotated gold-standard. Here, while we grouped the annotations into three categories: drugs, conditions/symptoms, and measurements. We did not use the systems' annotation categories, but rather their annotated terms and spans. This was done to accommodate the custom entity categories that systems like MedSpaCy and MedaCy have in their default settings and the fact that we are using only the first UMLS concepts identified by ScispaCy. Table 2 shows the annotation overlap analysis.

|  | Drugs | Conditions / symptoms | Measurements | Average |
| --- | --- | --- | --- | --- |
| **SMMT Tagger** | *69.31%* | *71.91%* | *39.83%* | **60.35%** |
| **MedSpaCy** | 19.98% | 13.49% | 7.45% | **13.64%** |
| **MedaCy** | 47.04% | 27.14% | 12.56% | **28.91%** |
| **ScispaCy** | 59.71% | 44.65% | 26.98% | **43.78%** |

Table 2. Annotation overlap analysis between gold standard dataset and evaluated systems.

We would like to stress again that MedSpaCy and MedaCy are at a disadvantage as their models are trained on considerably different data that does not work well with Twitter data. ScispaCy, however, performs fairly decently (in comparison) as the larger models capture relevant annotations when the tweet's text is clean and well-formed. It is out of the scope of this paper to properly tune these systems to ensure that they perform well with Twitter data, but it is certainly an interesting avenue for future research.

## Extraction evaluation on a limited set

While it is clear that regular text annotation performed the best in replicating the annotations that our clinicians made, we still annotated all 226,582,903 dataset tweets and evaluated the overlap of annotations made by the different systems. Table 3 shows the comparison between counts of produced annotations, processing time, and overlaps in annotations between the systems.

|  | Annotations Produced | Processing Time (minutes) | Overlaps with SMMT | Overlap with MedSpaCy | Overlap with MedaCy | Overlap with ScispaCy |
| --- | --- | --- | --- | --- | --- | --- |
| **SMMT Tagger** | 751,245,366 | **24,120** | **100%** | 20.12% | 33.91% | **72.28%** |
| **MedSpaCy** | 582,768,145 | 159,267 | 53.48% | **100%** | 42.23% | 55.39% |
| **MedaCy** | 656,311,799 | 26,147 | 51.14% | 44.92% | **100%** | 49.73% |
| **ScispaCy** | **775,615,621** | 325,620 | **89.17%** | 34.77% | 44.17% | **100%** |

Table 3. Annotation overlap evaluation for complete dataset.

# Conclusions

In this work we release a biomedically oriented automatically annotated dataset of COVID-19 chatter tweets. We demonstrate that while there are existing SpaCy-based systems for NER on clinical and scientific documents, they do not generalize well when used on non-clinical sources of data like tweets. However, we use this evaluation to justify the usage of a simple text tagger (SMMT) to produce annotations on a large set of tweets, based on its robustness when evaluated on a gold-standard manually curated dataset. The resulting dataset and biomedical annotations is the first and largest of its kind making it a substantial contribution with respect to using large-scale Twitter data for biomedical research. We have also added components for these types of tasks to SMMT, improving the usability of the resource.

As for future work, the release of this dataset will facilitate continued development of fine-tuned resources for mining social media data for biomedical and clinical applications. Recent research has shown social media data to be a valuable source of patient-reported information that is not available in similar granularity in other more traditional data sources.

# Conflicts of Interest

No potential conflict of interest relevant to this article was reported

# Acknowledgments

We would like to thank Jin-Dong Kim and the organizers of the virtual Biomedical Linked Annotation Hackathon 7 for providing us a space to work on this project and their valuable feedback during the online sessions. JMB funded by a grant by the National Institute on Aging (3P30AG059307-02S1).

## Authors' contribution

Conceptualization: JMB, TJC
Data curation: JMB, LR
Formal analysis: JMB, TJC
Methodology: JMB, LR, TJC
Writing – original draft: JMB, TJC, LR
Writing – review & editing: JMB, TJC, LR